# Ceramics Fragments Digitization by Photogrammetry, Reconstructions and Applications

J-B. Barreau, T. Nicolas, G. Bruniaux, E. Petit, Q. Petit,
Y. Bernard, R. Gaugne, V. Gouranton,
jean-baptiste.barreau@univ-rennes1.fr, theophane.nicolas@inrap.fr,
guillaumebruniaux1989@gmail.com, ptemilien@gmail.com, quentin.petit@irisa.fr,
yann.bernard@univ-rennes1.fr, ronan.gaugne@irisa.fr, valerie.gouranton@irisa.fr,

**Abstract**
This paper presents an application of photogrammetry on ceramic fragments from two excavation sites located north-west of France. The restitution by photogrammetry of these different fragments allowed reconstructions of the potteries in their original state or at least to get to as close as possible. We used the 3D reconstructions to compute some metrics and to generate a presentation support by using a 3D printer. This work is based on affordable tools and illustrates how 3D technologies can be quite easily integrated in archaeology process with limited financial resources.

**1. INTRODUCTION**

Today, photogrammetry and 3D modelling are an integral part of the methods used in archeology and heritage management. They provide answers to scientific needs in the fields of conservation, preservation, restoration and mediation of architectural, archaeological and cultural heritage [2] [6] [7] [9]. Photogrammetry on ceramic fragments was one of the first applications contemporary of the development of this technique applied in the archaeological community [3]. More recently and due to its democratization, it was applied more generally to artifacts [5]. Finally joined today by the rise of 3D printing [8] [10], it can restore fragmented artifacts [1] [12]. These examples target one or several particular objects and use different types of equipment that can be expensive. These aspects can put off uninitiated archaeologists. So it would be appropriate to see if these techniques could be generalized to a whole class of geometrically simple and common artifacts, such as ceramics.

From these observations, associated to ceramics specialists with fragments of broken ceramics, we aimed at arranging different tools and methods, including photogrammetry, to explore opportunities for a cheap and attainable reconstruction methodology and its possible applications. Our first objective was to establish a protocol for scanning fragments with photogrammetry, and for reconstruction of original ceramics. We used the digital reconstitutions of the ceramics we got following our process to calculate some metrics and to design and 3D print a display for the remaining fragments of one pottery.

1.1 Archaeological context
The archaeological material studied in this work consists in domestic ceramics from an Iron Age site in Rezé (Loire-Atlantique - France) and from a Bronze Age site in Lannion Penn An Alé (Côte d'Armor, France, excavated by S. Blanchet, Inrap).

**2. EQUIPMENT AND SOFTWARES**

2.1 Equipment
As the study is based on a reconstitution of archaeological ceramics with photogrammetry, it was essential to use a high quality digital single-lens reflex camera with a good lens and good image resolution. Indeed, this allows to obtain very good quality photos and better results during the search for common points with the photogrammetry software. We used a Nikon D60 Digital SLR to take photos of the various ceramic fragments [11]. Its technical features include a 10-megapixel CCD sensor, an 18-55mm optical AF-S with a 40mm focal distance and a Nikon autofocus module Multi-Cam 530 (3 AF points).

The environment used to take photos was composed of two light tables, a white background with its support and a halogen lamp (fig 1). This equipment allowed getting a light from below and above to minimize the shadow games. Homogeneity of brightness was difficult to obtain with only one single halogen lamp. This was a major problem because we needed a uniform and constant light regardless of the orientation of the shooting. Indeed, if the light intensity varies too much, then common points determined on a photo by photogrammetry software will not be the same from one photo to another.

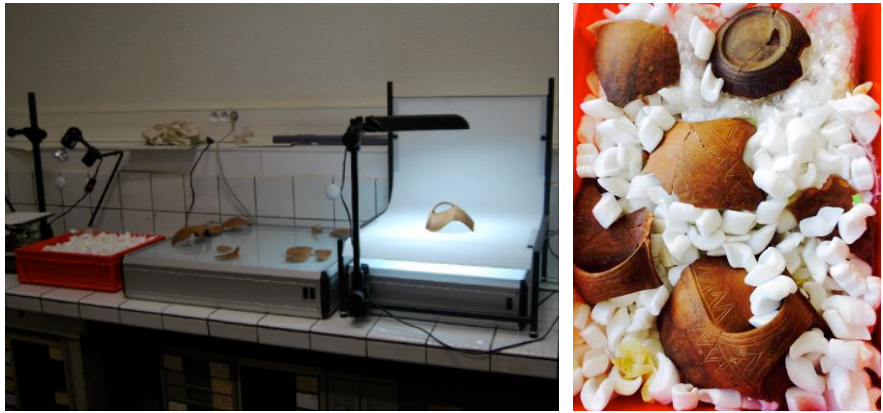

Figure 1. Shooting equipment and ceramic fragments

2.2 Software

For 3D restitution, we used the photogrammetry software Agisoft PhotoScan version 0.8.3 beta 64bits. This software has been chosen for its ease of use and its ability to generate highly detailed 3D models. 3D reconstruction from scanned fragments was performed via Blender software. The scale and volume calculation were processed with MeshLab software.

**3. 3D DIGITIZATION**

We present now the protocol followed to perform the 3D digitization of ceramic fragments by photogrammetry, the different steps to obtain realistic and satisfactory results, the technique used to calculate the volume of ceramics, and the problems we encountered during the work.

3.1 Shooting

The goal was here to shoot the object rotating on itself in order to get every viewing angle [4]. We took a first series of photos, describing a semicircle, with a distance camera-object as regular as possible. Then, we rotated the object through 180° in order to complete the series and to get a first coverage of the object. Other series of photographs were taken from different angles to cover a maximum area (Fig. 2).

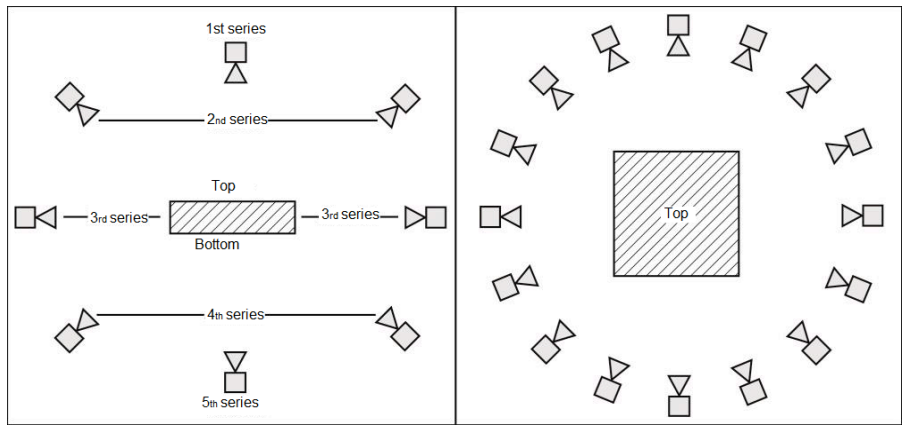

Figure 2. The different positions to take the series of photos. On the right, front view of the position of the series, and on the left, top view of the position of the camera in the 2$^{nd}$, 3$^{rd}$ and 4$^{th}$ series

3.2 Point cloud

Photos of the object were then imported into PhotoScan. Before starting the search for common points, a mask constraint (clipping of the object) must be included on each photo to focus the calculation only on the object and not all the environment. The search for common points and pictures alignment is then performed with a "strong" accuracy. Scaling the object is usually performed by placing two markers whose distance is known, on a photo of the object. The distance between the markers is set in Photoscan as a reference distance that enables the software to generate a 3D model with a correct scale. This functionality requires a sufficiently uniform light, which was not the case in our installation. According to this limitation, we chose to handle the scaling of the model in another further step. The maximum number of common points was set to 60000 instead of 40000 (default) to get a better rendering. The point cloud generated allowed to verify the coherency of the digital output with respect to the object (fig 3).

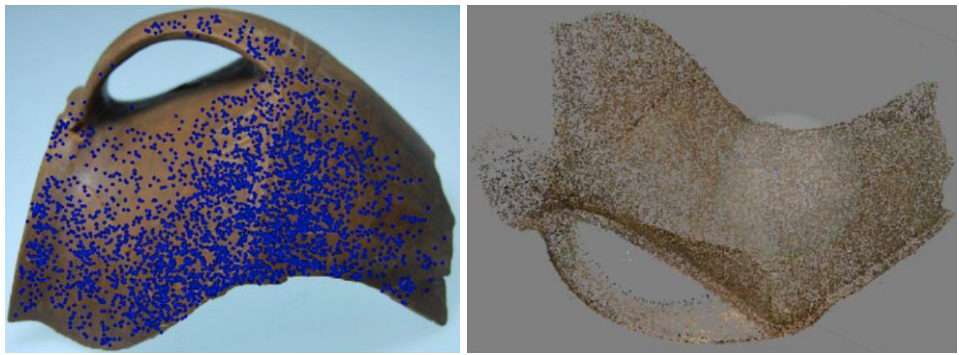

Figure 3. Piece of the Bronze Age ceramic with connection points (blue) determined by PhotoScan / point cloud get from the determination of connection points and photos alignment

3.3 Mesh and scale

The generation of geometries and textures was processed in PhotoScan with "arbitrary" option for the object type, "high" for the quality and "sharp" for geometric precision. The number of expected surfaces was set to 250000. The color of the mesh was computed from the color information contained in the points cloud. Photoscan computed the texture from several pieces of images extracted from the pictures, and applied it on the surface of the mesh (fig 4). The options for this step were a "generic" mapping mode and "average" fusion mode.

The resulting mesh was then exported in the ".obj" format and imported in Meshlab to perform the scaling. A measure of distance was computed with the tool "rule" between two points whose real distance is known. The ratio between the real distance and the measured distance on the mesh provided the proportionality factor between these two distances. Thanks to the "Transform: scale" MeshLab function, distances on the X, Y and Z axis were multiplied by this factor achieving the correct scaling of the mesh.

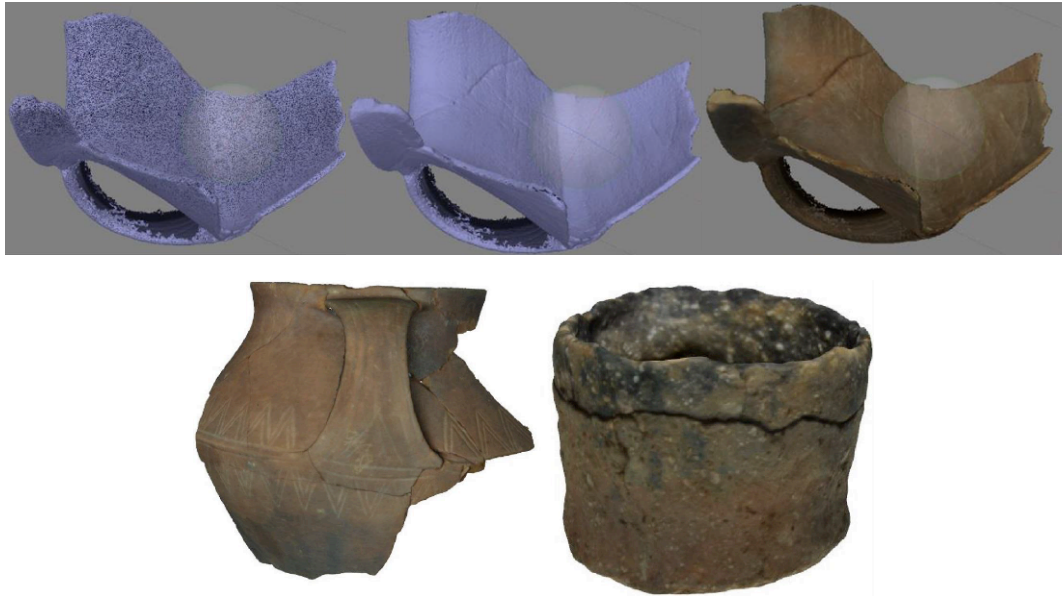

Figure 4. Raw mesh (top left) / smoothed mesh (top middle) / mesh with colors (top right) / textured meshs (bottom)

## 4. RECONSTRUCTIONS AND APPLICATIONS

### 4.1 Rezé ceramics

In the case of the Rezé ceramics, the available fragments cover only a partial part of each object. Therefore, the reconstruction was theoretical and performed by extrapolating the different shapes from the fragments in Blender. Most of the fragments contain a part of the lip, the neck, the edge, the body or the bottom. Therefore, if the fragment has a piece of ceramic bottom, it can be aligned with the horizontal plane XY for an easier handling. The lip of the ceramic fragment has a curvature. Using this curvature, a circle (circular mesh) was created and the perimeter of this circle adjusted to coincide perfectly with the contour of the lip. This provided the theoretical circumference of the upper ceramic part. The same operation was repeated several times by shifting the other circles on the Z-axis which were then adjusted on the other parts of the ceramic (kept and visible on the fragment). Therefore, the assembly of these circles on the fragment provided the "skeleton" of the ceramic. Once this frame done, it was imported in MeshLab again to calculate the volume of the ceramic. The "convex hull" function, in the "remeshing, simplification and reconstruction" tab allowed to create a mesh from the different circles of the frame (fig 5). This mesh, constituting a completely closed shape, was used to calculate a volume with the "compute geometric measures" function. The table 1 gathers the characteristics of the reconstitutions of the different Rezé ceramics.

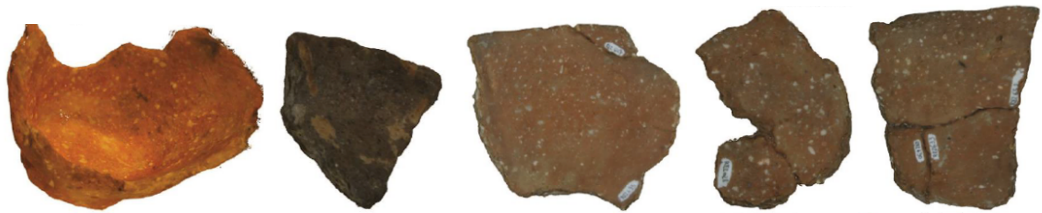

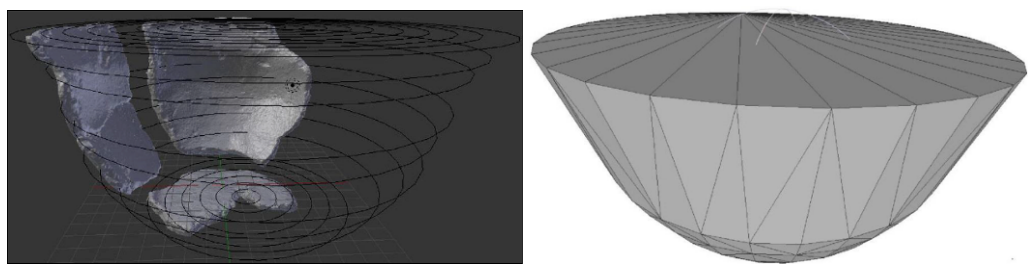

Figure 5. Fragments of eating ceramics reconstituted by photogrammetry / Frame of circles aligned along the Z axis and dimensioned according to the curvature of the fragments / "convex hull" function on the Gr8B ceramic

Table 1. Summary of the characteristics of each reconstituted ceramic fragments

| Ceramic name | Number of photos | Number of points (point cloud) | Number of surfaces | Number of vertices | Calculation time | Volume (cm3) |
|---|---|---|---|---|---|---|
| Gr1C | 36 | 24101 | 249 999 | 130234 | 40m | 536,4 |
| Gr5E | 78 | 55470 | 246768 | 128077 | 1h14m | 1722,8 |
| MGr2E | 89 | 31204 | 249999 | 139674 | 52m | 88,9 |
| MGr11D | 75 | 18354 | 250000 | 129272 | 43m | 138,8 |
| Gr1E | 44 | 29396 | 17911 | 11036 | 39m | 338,7 |
| Gr7D | 62 | 45250 | 154286 | 77517 | 41m | 211,1 |
| Gr8B | 111 | 59623 | 250000 | 149449 | 1h15m | 1938,7 |

4.2 Lannion Penn An Alé ceramic

This Bronze Age ceramic is "almost" full, so we tried to assemble the different fragments in Blender. Fragments of ceramic having a part of the lip were first assembled and wedged on a circle perfectly matching its shape. The other pieces were then imported and positioned one by one. One difficulty, however, was met with several small fragments. For them, only the inner or outer part of the fragment was reconstituted. The reasons for this failure are many and difficult to determine. Indeed, this can be induced by low quality of light, insufficient image quality or a too thin thickness of the ceramic fragments that corrupted the detection of common points on the edges. To solve this problem, the inner and outer parts were rescanned separately and then scaled in MeshLab. The two parts were then positioned parallel to the horizontal plane XY, superimposed on each other by playing with a rotation on the Z axis, and then merged by sliding the inner part relative to the Z axis.

The reconstruction approach was here different. After deleting the handles, we used the function "Surface reconstruction: Poisson" in Meshlab allowing to extrapolate the initial ceramic shape from the assembly of fragments (fig 6). This shape was used to compute the volume of the ceramic whose result is 2742,9 cm3.

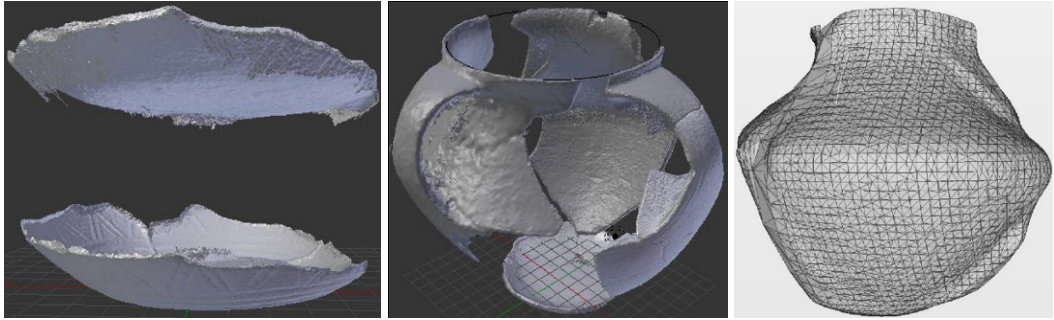

Figure 6. Inner part (top) and outer part (bottom) of the bottom / reconstruction step / reconstruction with the function "Surface reconstruction: Poisson"

A last step of the work consisted in using the generated shape to build a support for the fragments, in order to get a 3D tangible representation of the pottery. We used a 3D printer MarkerBot Replicator 2x, which is an ABS printer, to produce the support display. The mesh required some corrections made by a graphic designer under 3DSmax (fig 7), because the Poisson reconstruction introduced some approximation at the frontiers between the real fragments and the computed shape.

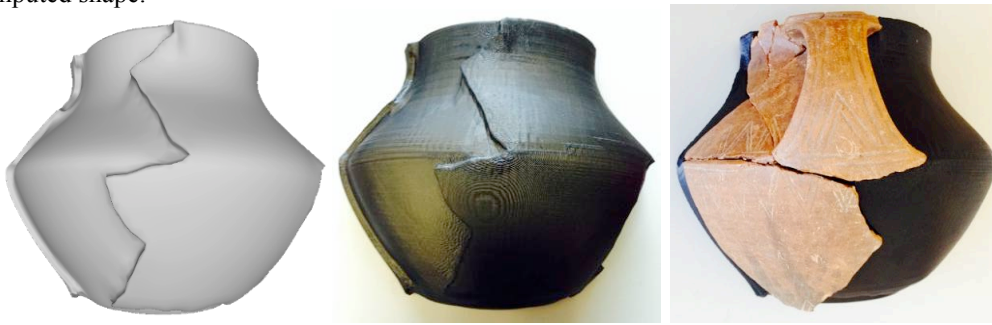

Figure 7. 3D model of the display/ empty display / display with fragments

## 5. DISCUSSION

The digitization process has highlighted difficulties often masked by the apparent simplicity of photogrammetry. However, we believe that the goal was reached regarding the digitization of fragments, and for a very low cost. The different approaches about reconstruction discussed here provided practical useful data for the analysis work performed by archaeologists. The estimate of their volumes, the 3D visualization of the shape, and the tangible 3D presentation represented concrete help in the study of the ceramics.

The 3D printer used for this work is a relatively inexpensive one, recently bought by the computer science laboratory involved in the project. The generation of a support display for the Lannion ceramic was a proof of concept for presenting fragments physically held around the content. However, problems have arisen. Because the Bronze Age ceramic is too large for this printer, especially because of the size of the handles, we had to print the support in two parts. In addition, the Poisson reconstruction of the internal volume of the ceramic generated approximations, which implied that the fragments did not exactly fit the display. Modifications on the 3d model of the display to match the shape of the fragments were required.

The result of this experimental work convinced us that the production such display could well be generalized within the museum community due to its low cost. A first reflection on its use has led us to believe that it would be wise to add on top of the display (not covered by a fragment) signs indicating metadata in the print-ready 3D model (fig 8).

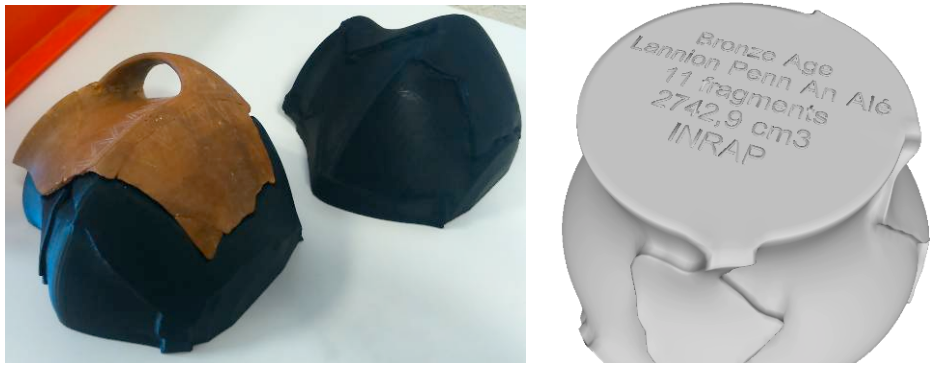

Figure 8. Separated part of the display / metadata inscription on the 3d model

## 6. CONCLUSION AND FUTURE WORKS

We proposed a simple protocol to digitize fragments of ceramics. This protocol is accurate enough to provide a digital reconstitution of ceramics with heterogeneous morphologies and compute basic metrics. Furthermore, from the digital reconstitution of the Rezé ceramics, we designed and 3D printed a display for the remaining fragments. This kind of display can be used for exhibition of artefacts in museums.

The next step of this work is to validate these methods on more ceramics. To do this, we must solve the problems of accuracy in digitization and printing of large volumes. We also wish to extend these methods to other kinds of less symmetrical archaeological fragmented artifacts. Finally, it would be interesting to integrate museologists to our future works on 3D printed displays.